\def\HA{{H$\alpha$}}

\def\kms{\:\rm{\,km\,s^{-1}}}

\def\asy{\rm{\arcsec\,yr^{-1}}}

\def\OiL{[\ion{O}{3}] $\lambda 6300$}

\def\OiiiL{[\ion{O}{3}] $\lambda\lambda 4959,5007$}
\def\SiiL{[\ion{S}{2}] $\lambda\lambda 6717, 6731$}

\def\OiL{[\ion{O}{1}] $\lambda\lambda 6300, 6363$}

\def\sii{[\ion{S}{2}]}

\def\oiii{[\ion{O}{3}]}

\def\hi{\ion{H}{1}}

\documentclass[12pt,preprint]{aastex}

\begin{document}


\newcommand{\MSOL}{\mbox{$\:M_{\sun}$}}  

\newcommand{\EXPN}[2]{\mbox{$#1\times 10^{#2}$}}
\newcommand{\EXPU}[3]{\mbox{\rm $#1 \times 10^{#2} \rm\:#3$}}  
\newcommand{\POW}[2]{\mbox{$\rm10^{#1}\rm\:#2$}}
\newcommand{\SING}[2]{#1$\thinspace \lambda $#2}
\newcommand{\MULT}[2]{#1$\thinspace \lambda \lambda $#2}
\newcommand{\CHINU}{\mbox{$\chi_{\nu}^2$}}
\newcommand{\vsini}{\mbox{$v\:\sin{(i)}$}}

\newcommand{\fuse}{{\it FUSE}}
\newcommand{\hst}{{\it HST}}
\newcommand{\iue}{{\it IUE}}
\newcommand{\euve}{{\it EUVE}}
\newcommand{\einstein}{{\it Einstein}}
\newcommand{\rosat}{{\it ROSAT}}
\newcommand{\chandra}{{\it Chandra}}
\newcommand{\xmm}{{\it XMM-Newton}}
\newcommand{\swift}{{\it Swift}}
\newcommand{\asca}{{\it ASCA}}
\newcommand{\galex}{{\it GALEX}}
\newcommand{\cxo}{CXO\,1337}

\newcommand{\msun}{M_\odot}
\newcommand{\src}{G1.9+0.3}
\newcommand{\tbn}{$\theta_{\rm Bn}$}
\newcommand{\roll}{$\nu_{\rm roll}$}
\def\etal{{et~al.}}

\catcode`\@=11
\newcommand{\gapprox}{\mathrel{\mathpalette\@versim>}}
\newcommand{\lapprox}{\mathrel{\mathpalette\@versim<}}
\newcommand{\propapprox}{\mathrel{\mathpalette\@versim\propto}}
\newcommand{\@versim}[2]
  {\lower3.1truept\vbox{\baselineskip0pt\lineskip0.5truept
\ialign{$\m@th#1\hfil##\hfil$\crcr#2\crcr\sim\crcr}}}
\catcode`\@=12


\shorttitle{X-ray Proper Motions on NW Rim of SN\,1006}
\shortauthors{Katsuda \etal}

\slugcomment{Accepted for publication in {\em The Astrophysical Journal}}
{\title{X-ray Proper Motions and Shock Speeds along the 
 Northwest Rim of SN\,1006\footnote{
Based on observations made with NASA's \chandra\ X-ray Observatory, 
NASA's \chandra\ Observatory is operated by Smithsonian Astrophysical Observatory 
under contract \# NAS83060 and the data were obtained through program GO1-12115.}}

\author{
Satoru Katsuda\altaffilmark{2},
Knox S. Long\altaffilmark{3},
Robert Petre\altaffilmark{4},
Stephen P. Reynolds\altaffilmark{5},
Brian J. Williams\altaffilmark{3}, and
P. Frank Winkler\altaffilmark{6}
}
\altaffiltext{2}{RIKEN (The Institute of Physical and Chemical Research), 2-1 Hirosawa, Wako, Saitama 351-0198, Japan; katsuda@crab.riken.jp}
\altaffiltext{3}{Space Telescope Science Institute, 3700 San Martin Drive, Baltimore, MD, 21218;  long@stsci.edu} 
\altaffiltext{4}{NASA Goddard Space Flight Center, Greenbelt, MD 20771; brian.j.williams@nasa.gov}
\altaffiltext{5}{Physics Department, North Carolina State University, Raleigh, NC 27695; reynolds@ncsu.edu}
\altaffiltext{6}{Department of Physics, Middlebury College, Middlebury, VT, 05753; 
winkler@middlebury.edu}

\begin{abstract}

We report the results of an X-ray proper motion measurement for the NW rim of SN\,1006, carried out by comparing  \chandra\ observations from 2001 and 2012.  The NW limb has predominantly thermal X-ray emission, and it is the only location in SN\,1006 with significant optical emission: a thin, Balmer-dominated filament.  For most of the NW rim, the proper motion is  $\approx 0.30 \asy$, essentially the same as has been measured from the \HA\ filament. 
Isolated  regions of the NW limb are dominated by nonthermal emission, and here the  proper motion is much higher, $0.49 \asy$, close to the value measured in  X-rays along the much brighter NE limb, where the X-rays are overwhelmingly  nonthermal.  At the 2.2 kpc distance to SN\,1006, the proper motions imply shock velocities of $\sim 3000$ km\,s$^{-1}$ and $\sim 5000$ km\,s$^{-1}$ in the thermal and nonthermal regions, respectively.  A lower velocity behind the \HA\ filament is consistent with the picture  that  SN\,1006  is encountering denser gas in the NW, as is also suggested by its overall morphology.   In the thermally-dominated portion of the X-ray shell, we also see an offset in the radial profiles at different energies; the  0.5--0.6\,keV peak dominated by \ion{O}{7} is closer to  the shock front than that of the 0.8--3\,keV emission---due to the longer times for heavier elements to reach ionization states where they produce strong X-ray emission. 
 
\end{abstract}



\section{Introduction \label{sec_intro}}
\label{intro}

The supernova of SN\,1006 was observed widely in Asia and Europe, and
appears to have been the brightest naked-eye supernova ever recorded
\citep{stephenson10}.  Its remnant appears in catalog listings   as SNR G327.6+14.6, though  in this paper we  adopt the common parlance of using ``SN\,1006" to refer to the remnant.  SN\,1006
stands out among historical remnants as  largest (30\arcmin = 19 pc diameter), one of the closest, at 2.2 kpc \citep{winkler03}, farthest above the Galactic plane (550 pc), and least obscured ($N_H \approx 5\times10^{20}\;{\rm cm^{-2}}$, all of which make it  amenable to study in X-rays and other bands.   However, it is also the faintest of the historical remnants, in total flux and especially in surface brightness, so deep observations are required for detailed study.  Since SN 1006 is both a prototypical Type Ia
SNR, and since it shows some of the best and cleanest evidence for
electron acceleration to TeV energies in shocks, its close examination
can further our understanding of both SNR dynamics and particle
acceleration in general.

The overall  structure of SN\,1006 is that of a nearly circular limb-brightened shell with strong bilateral symmetry; its  NE and SW limbs  are far brighter, in both radio and X-rays, than other portions of the rim \citep[][and references therein]{cassam-chenai08, dyer09, miceli09}.  \citet{koyama95} found that the X-rays from the bright NE and SW limbs are not only harder than those from the rest of SN\,1006, but also that their spectrum is a featureless power-law, a result that provided the first clear evidence for diffusive acceleration of charged particles and cemented the long-suspected link between supernova remnant (SNR) shocks and cosmic rays \citep{reynolds96}.  Furthermore,  the X-ray morphology along  the NE and SW rims matches the radio images in exquisite detail \citep{winkler97}, further confirming a common synchrotron origin for emission in both widely separated bands.   

The NW and SE limbs of SN\,1006 are dramatically different; the shell in both the NW and SE is much   fainter and less distinct in the X-ray and  radio bands, and the X-ray spectrum is soft and thermal, dominated by lines from He-like and H-like oxygen between 0.5 and 0.8 keV \citep{long03}.  The NW limb is unique in that only here can significant optical emission  be seen: a set of delicate filaments along the outer NW rim, with emission that consists solely of hydrogen Balmer lines \citep{vandenbergh76,schweizer78,ghavamian02}.\footnote{In addition to the NW filaments, there is exceedingly faint and more diffuse Balmer emission surrounding almost all of the SN\,1006 shell \citep{winkler97,winkler03}.}   First observed in the remnant of Tycho's supernova \citep{kirshner78}, Balmer-dominated, or ``nonradiative" filaments arise when a fast shock expands  
 into a low-density, partially neutral environment.  
The Balmer emission has two components:  neutral H atoms that enter the shock  can
be excited, and subsequently decay to produce narrow lines (with a width
characteristic of the pre-shock temperature); or they can undergo
charge exchange with hot protons to produce broad lines whose width is
closely related to the post-shock proton temperature \citep[e.g.,][]
{chevalier80, ghavamian02, heng10}.  Since the lifetime of neutral atoms in the hot
post-shock environment is very short, the Balmer filaments can occur
only {\em immediately} behind the shock, and thus delineate the
current position of the shock.

Proper motions of the brightest Balmer filaments in the NW were first measured by \citet{hesser81} using photographic plates taken five years apart, and more recently by \citet{long88} and by \citet{winkler03}.  The latter measurement, using CCD images taken over an 11-year baseline, obtained $\mu_{{\rm H}\alpha} = 0.280 \pm 0.008 \asy$. 
Based on the Balmer profiles and models for the shock conditions, \citet{ghavamian02} obtained a value of $v_s = 2890 \pm 100 \kms$ for the shock velocity in the NW, and combining this value with the proper-motion measurement yields a geometrically determined distance of $2.2 \pm 0.1$ kpc to SN\,1006.  Having a relatively precise distance of course means that proper motions measured by any means, anywhere around the SN\,1006 shell, translate directly into a shock velocity at that point, which is significant because $v_s$ is the most crucial parameter in the theory of shock acceleration \citep[e.g.,][]{reynolds08}.   

A useful way of expressing the proper motion is through the expansion parameter $m$: the power-law index in $R \propto t^m$, where $t$ is the age of the remnant.  This parameter can be interpreted as the ratio of current expansion rate divided by the mean rate over the remnant's lifetime, $m=\mu t/\theta$, where $\theta$ is the angular radius.   The expansion index gives some insight into a remnant's evolutionary state, e.g., $m = 1$ for free expansion, or $m=0.4$ for  Sedov expansion.

Proper-motion measurements of SN\,1006 have  been made in  
various bands with different precision.
Radio measurements of the {\em global} expansion gave 
$m_{{\rm radio}} = 0.48\pm 0.13$
\citep{moffett93}, considerably larger than the value of $m_{{\rm H}\alpha} = 0.34 \pm 0.01$ for 
the NW optical filament alone \citep{winkler03}.  Since the stronger
optical emission in the NW indicates a higher external density, it is not
surprising that the expansion should be slower than the average around the shell, since the higher the density, the more rapidly   the expansion will decelerate.  (If one assumes pressure equilibrium, which may pertain in SN\,1006 but is not assured, the preshock density will scale inversely as the square root of the shock velocity.)
In a recent paper, we obtained an  X-ray measurement of the proper motions along the E-NE limb of SN\,1006, using a pair of \chandra\ ACIS-S observations from 2000 and 2008, and found $\mu = 0.48 \pm 0.04 \asy$, or $m = 0.54 \pm 0.05$, and  $v_s = 5000 \pm 400 \kms$ \citep{katsuda09}.

In this paper, we present the first X-ray measurement of the proper motion along the NW limb of SN\,1006, using a pair of deep \chandra\ ACIS-S observations from 2001 and  2012.  These new measurements cover much the same region as the optical proper motion measurement by \citet{winkler03}, and the results are consistent.   In addition to the thermal X-ray emission that dominates most of the NW limb, there are two small regions with a much harder  X-ray spectrum, free of emission lines and consistent with nonthermal synchrotron emission.  We have also measured the motions in these regions and find these to be consistent with our earlier measurements \citep{katsuda09} for the much brighter synchrotron-dominated NE rim.   The 2012 April observation of the NW limb represents the first in a series of deep ACIS images covering the entirety SN\,1006, being carried out as a \chandra\ Large Project.   Further results will be reported in subsequent publications.

\section{Observations and Results}

The initial \chandra\ observation of the NW limb was carried out with the ACIS-S in 2001 April and reported by \citet{long03}.  In order to measure proper motions in the NW, we carried out a repeat observation in 2012 April; the roll angle and exposure time were 
almost identical for both observations, and the positions were similar, 
with the second-epoch one displaced slightly to the NW 
in order to include more of the pre-shock background.   The two observations  are summarized in 
Table~\ref{tab:obsns}.  In both cases we reprocessed 
the level-1 event files with CIAO ver. 4.4 and CALDB ver.4.5.1.

A three-color image of a portion of the NW limb of SN\,1006 as seen with 
\chandra\ in 2012 is shown in Figure~\ref{fig:image}, where red, green, 
and blue correspond to 0.5--1\,keV, 1--2\,keV, and 2--7\,keV bands, respectively.  
Most of the X-ray emission in the NW is thermal, and appears reddish in this image, but there are also a few fainter nonthermal, synchrotron-dominated regions that appear more bluish.  The difference in spectral character between the thermal and nonthermal regions is most apparent in spectra from these regions, shown in Fig.~\ref{fig:spectra}.

For the sharp, bright shock front in the NE, an outward displacement was evident in a simple difference image between the images at epochs eight years apart \citep{katsuda09}.  In the NW, however, the surface brightness is much lower and the shock front less sharp, so the expansion is less obvious.  In order to measure it, we have integrated along the shock front to give radial profiles in three thermal regions where the shock is well defined, labeled T1, T2, and T3, and also in two nonthermal regions, labeled NT1 and NT2.  

As described in Sections 2.2 and 2.3, we have measured the proper motions in different regions in different energy bands.  In order to produce flux-calibrated images, it was necessary  to consider the variation in effective exposure time with both energy and position on the detector.   We  created  exposure maps  at monochromatic energies: 0.6\,keV, 0.55\,keV, 1.2\,keV, and 2\,keV for 0.5--1\,keV, 0.5--0.6\,keV, 0.8--3\,keV, and 1--8\,keV bands, respectively---energies that roughly correspond to mean photon energies in the energy bands of interest, given the spectral shape in each region (Fig.~\ref{fig:spectra}).

\subsection{Image Registration}
Before measuring the proper motion, we first registered the images using as fiducials four bright point sources, all 
with significance levels greater than $10\,\sigma$ and point-spread-function (PSF) 
sizes less than 2\arcsec\ in both images.\footnote{Here PSF size is defined as a radius in which 
39\% of total source photons are enclosed.}
We used the CIAO tool {\tt wavdetect} to obtain positions 
for the sources  at the two epochs,  which we summarize in Table~\ref{tab:points}; the sources are also indicated in the cyan circles on Fig.~\ref{fig:image}. 
 Sources P1 and P3 correspond to faint ($R \approx 19$) star-like objects in the NOMAD catalog, \citep{zacharias05}, both with negligible proper motion, while the P2 and P4 sources seem to have no optical companion.\footnote{Source P2 is located 2.5\arcsec\ NW of another faint object in the NOMAD catalog, but since the position difference is well outside the error ellipses from both ACIS observations, this is unlikely to be a true association.  In any case, this object also has no proper motion in the NOMAD catalog.}  All four sources have a hard spectrum, and all are likely to be background AGNs; we assume that all have negligible proper motions.  
The offset between the two images, obtained from a simple mean of the position differences listed in the two right-hand columns of Table~\ref{tab:points},  are 0.06\arcsec\ and 0.18\arcsec, in R.A.\ and Decl., respectively.  Such offsets   are consistent with the stated absolute astrometric 
accuracy for {\em Chandra} and  the ACIS-S array.\footnote{http://cxc.harvard.edu/cal/ASPECT/celmon/.}
After correcting for these offsets, the RMS residuals in position difference between the two epochs for the four point sources   are 0.37\arcsec\ in R.A. and 0.27\arcsec\ in Decl.  In our proper motion analysis that follows, we have applied the mean offsets, and have included both statistical uncertainty and uncertainty  in the registration (for the radial direction),  calculated  as $\sqrt {{{(0.37''\sin \theta )}^2} + {{(0.27''\cos \theta )}^2}}$, where $\theta$ is the azimuth angle for each area, measured counterclockwise from North.

\subsection{Thermal Regions}
We selected regions T1, T2, and T3 both because they coincide with regions studied previously and because the shock front is sharply defined there.  
Region T1  includes a small, bright X-ray knot,  located within an \HA\ bubble ahead of most of the NW shock, and also includes the bright X-ray knot where \citet{vink03} directly measured both electron and ion temperatures  using {\em XMM-Newton} reflection grating spectra.   They demonstrated that $T_e$ remains much lower than $T_i$ well behind the shock, and also that non-equilibrium ionization conditions prevail.
Regions T2 and T3  trace the shock front  delineating 
the NW limb where  proper motions of the \HA\ filaments have been precisely measured \citep{long88,winkler03}.  

In Figure~\ref{fig:rad_profT}, we plot the  projected one-dimensional profiles 
for epochs 2001 and 2012, which 
clearly show outward motion in all three regions.  The profiles have been generated from 
vignetting-corrected images in the 0.5--1\,keV band,  which contains a 
large fraction of the total X-ray emission.    
To quantitatively measure the shifts between epochs, we have followed the approach we  used 
previously \citep[e.g.,][]{winkler03,katsuda09}.  Briefly, the method is to minimize the $\chi^{2}$ value for the difference between the second-epoch profile and the shifted first-epoch one, as a function of the amount of the shift.  
When calculating the $\chi^2$ values, we use data points within the 
vertical dashed lines indicated in each panel of 
Figure~\ref{fig:rad_profT}, so that we can concentrate on features of interest.  The results of these measurements for all three regions are given in Table~\ref{tab:prop}.  
 
The X-ray proper 
motion we measure in region T2 is  consistent, within the uncertainty, with the optical measurement
for the same region \citep[$0.280\arcsec \pm 0.008 \asy$:][]{winkler03}.
The proper motions in regions T1 and T3 are slightly higher than in T2, but all are consistent within the uncertainties.   Taking the distance to SN\,1006 as 2.2 kpc \citep{winkler03}, the expansion velocity for T1 is $3300 \pm 200  \pm 300\kms$ (statistical and registration uncertainties, respectively), in reasonable agreement with the velocity inferred from the X-ray line width, where \citet{vink03} found that the  velocity  is $\gtrsim 4000 \kms$, if no significant temperature equilibration has taken place, but could be as low as $\sim 3000 \kms$ with some adiabatic cooling.

The first radio measurement of the SN\,1006 expansion was by \citet{moffett93}, who measured a global average expansion rate of $0.44\arcsec \pm 0.13 \asy$ over a baseline of only 8 years.  They also give local rates in four broad azimuthal sectors, but the NW rim was too indistinct to give any measurement at all there.  However, \citet{moffett04} have since obtained a third-epoch radio image of SN\,1006, and with  a longer baseline obtained expansion rates in the NW that are consistent with the optical values. 

In Table~\ref{tab:prop}, we also give the local values for  the expansion index, $m$.  Calculating $m = \mu t/\theta$ requires an angular radius, $\theta$, and to define this we have 
taken the expansion center to be R.A. (J2000.)\ = 
15:02:54.9, Decl. (J2000.)\ = -41:56:08.9, as determined by \citet{katsuda09} from the 
{\it ROSAT} HRI X-ray image mosaic of  the entire remnant. 

For Area T2, we have also examined radial profiles and proper motions in distinct energy 
bands.   Of the three thermal regions, T2 is the best suited  for such a study because it has the 
sharpest shock front and the best photon statistics.  As shown in 
Figure~\ref{fig:rad_profT2} (left), the radial profiles vary with X-ray 
energy; the peak of the 0.5--0.6\ keV (a band dominated by \ion{O}{7} triplets) profile 
is located  just behind the H$\alpha$ filament, whereas the  0.8--3\ keV peak
is broader and shifted by $\sim 15\arcsec$ further behind the shock.
This  must be primarily due to the evolution of the ionization to produce He-like species of elements heavier than oxygen, as shown in the profiles and spectra along the NW rim that appear in \citet{long03}.  
Ionization of higher-Z metals, as well as the gradual increase in the  electron temperature behind the shock, both make  the X-ray spectrum  harder toward 
the center of the SNR.  We have measured proper motions in these two 
energy bands, using the one-dimensional profiles shown in 
Figure~\ref{fig:rad_profT2} (center and right), and find no significant energy dependence; the values are given in Table~\ref{tab:prop}.

\subsection{Nonthermal Regions}

We have also carried out proper-motion measurements for the small regions in the NW that are dominated by nonthermal X-rays, 
i.e., areas NT1 and NT2 in Figure~\ref{fig:image}.  Since the X-ray emission in these areas
is harder, we have used a higher and broader energy band, 1--8\ keV, 
 to measure the proper motion.  The radial profiles in 
Figure~\ref{fig:rad_profNT} show clear outward motions that are significantly larger than those found for the thermal regions.
Measurements using the same $\chi ^2$ minimization technique  yield almost identical values of $0.48\; {\rm to}\;0.49 \asy$ for both NT1 and NT2, both  $\sim 50\%$ higher than the values obtained in the thermal regions.  Furthermore, the proper motions in NW regions NT1 and NT2 are consistent with our previous measurements of $0.48 \pm 0.04 \asy$ measured at more than a dozen locations along the shock in  the much brighter synchrotron-dominated NE region of SN\,1006 \citep{katsuda09}.

\section{Discussion}

We have confirmed that for the thermal-dominated NW shock, the proper motions measured in X-rays are consistent with those measured optically for the Balmer-dominated filaments that delineate the shock.  While not unexpected, this is the first time that proper motions have been measured in both X-ray and optical bands for the same region in SN\,1006.   This is noteworthy because for other young remnants, past measurements in different bands have sometimes produced widely discrepant  results \citep[e.g., see the summary in][]{katsuda09}, though these may well have been due to comparing local measurements with global averages.  
For the Tycho SNR, at least, recent high-resolution measurements with \chandra\  by \citet{katsuda10} are in reasonable  agreement with radio measurements by \citet{reynoso97} around the entire shell.  We have now seen that for SN\,1006, X-ray and optical measurements along the NW rim are in excellent agreement.   
 But as the capability for  detailed measurements becomes available, it is important to keep in mind that the features themselves can evolve, and that new features can appear while others disappear, giving the illusion of proper motions in snapshot images taken many years apart.   

Our confirmation that the shock velocity in the thermal-dominated NW limb is slower
than at other locations around the SNR shell provides direct evidence that the SNR shock
is interacting with a denser ambient medium in the NW region, as has previously been
suggested by several studies. \citet{heng07} found that the Balmer-dominated H$\alpha$
filaments in the NW \citep[which have  significantly higher surface brightness than optical emission elsewhere 
in SN\,1006,][]{winkler03} are consistent with a pre-shock density of 0.15-0.3 cm$^{-3}$, higher by
a factor of several from that around the rest of the remnant. This density is similar to
that inferred from X-ray spectra from {\it Chandra} \citep{long03} and {\it XMM-Newton}
\citep{acero07,miceli09}, as well as from {\it FUSE}  UV spectra  \citep{korreck04}. Most recently,
\citet{winkler12} report the detection of 24$\mu$m infrared emission arising from warm dust grains in
SN\,1006. This emission is  seen only along the NW rim, and {\em Spitzer} spectra of the warm dust
is consistent with a post-shock density of 1\,cm$^{-3}$.

The shock in the NW of SN 1006 is apparently encountering a localized region of higher density than
that surrounding the rest of the remnant, possibly representing the start of an encounter with the extended \hi\ concentration 
just beyond the rim found by \citet{dubner02}.  As expected where the pre-shock density is higher than average, the rim in the NW is indeed flatter than
elsewhere,  with a radius up to  15\% smaller there than the average around SN\,1006. Since the NW filament
lies at a radius only slightly smaller than the remnant mean, this slower expansion cannot have been
going on for very long, and so the shock must have encountered the denser region fairly recently.

Another point of view is to consider the expansion index $m$. In the E and NE, $m \approx 0.54$, indicating that the
expansion is well short of reaching the Sedov phase ($m = 0.4$) in these regions, consistent with expansion
into a very low-density environment \citep{katsuda09}. But in the NW, where the radius is only slightly smaller than elsewhere, $m \approx 0.38$, consistent with Sedov expansion, and thus with a higher density.

The small sections of the NW limb where nonthermal  X-ray emission 
dominates have proper motions, and hence shock velocities,  significantly higher 
 than those in the thermal regions.  
As noted above, the measured values of  $\sim0.5 \asy$ for the nonthermal regions NT1 and NT2 give  shock velocities  $\sim 5000\kms $ at a distance of 2.2\,kpc, similar to those all along  the 
nonthermal, synchrotron-dominated NE limb \citep{katsuda09}.  Thus it 
appears likely that in SN\,1006, nonthermal-dominated regions are commonly associated with 
fast shocks of $\sim 5000\kms $.   Higher shock speeds than in the thermal regions indicate a  lower  pre-shock density for the the nonthermal regions, suggesting the existence of low-density pockets within the generally  higher ISM density to the NW, or other small scale inhomogeneities.  

Higher shock speeds in the nonthermal regions  are also consistent 
 with the theoretical view that 
faster shocks can enhance synchrotron X-ray emission by boosting the 
roll-off frequency $\nu_{\rm rolloff}$.  Diffusive shock acceleration theory predicts a power-law electron energy distribution with an exponential cutoff at some characteristic energy, and $\nu_{\rm rolloff}$ is the peak frequency emitted by electrons with that energy \citep[e.g.,][]{ellison00}. 
\citet{reynolds08} has shown that 
$\nu_{\rm rolloff} \propto v_{\rm s}^{4}$ when the maximum energy of 
accelerated particles is limited by the SNR age, or 
$\nu_{\rm rolloff} \propto v_{\rm s}^{2}$  when synchrotron losses limit the maximum energy.

Still,  a shock speed of $\sim 3000 \kms$ should be high enough for substantial
particle acceleration to X-ray-emitting energies.  The strength and orientation of the upstream magnetic field, as well as the  electron diffusion coefficient, also play important roles in 
determining acceleration efficiency, and could inhibit particle acceleration in the NW while enhancing it in the NE and SW.  
 In addition, a partially neutral pre-shock medium, a condition that the existence of bright Balmer filaments indicate must pertain along most of the NW limb,  may inhibit 
shock acceleration  \citep{raymond11, blasi12}. 
In particular, Blasi et al. find that for shock speeds below $3000 \kms$, the energy spectrum for accelerated particles can steepen markedly at low energies, reducing their X-ray synchrotron emission compared to that from faster shocks.

Finally, we note that the SE portion of the remnant has predominantly thermal X-ray emission \citep{rothenflug04}, but resembles the nonthermal regions in terms of the observed local remnant radius (which implies the mean shock velocity over the lifetime of the remnant) and in the faintness of its \HA\ emission (which implies the ambient density).  \citet{cassam-chenai08} have attributed the dramatic difference largely to a far lower efficiency for particle acceleration to the SE than in the NE or SW\@.  Measuring proper motions along the SE rim should be an important diagnostic for  understanding the nature of particle acceleration in SN\,1006.

\section{Summary}

We have measured X-ray proper motions in the NW limb of 
SN\,1006, using the first of our new \chandra\ Large Project observations.
For the thermally-dominated  X-ray regions along most of the NW limb, we find proper motions of $ 0.28$--$0.35\asy$, consistent (within the uncertainties) with optical measurements for the Balmer filaments, $ 0.28\asy $,  by \citet{winkler03}.
Even along the thermal-dominated NW rim, however, we find two small,  isolated regions with harder, non-thermal X-ray emission.   Proper motion measurements for these regions are much higher: $\mu_X \approx 0.49 \asy$, almost identical to the values measured by \citet{katsuda09} along  the synchrotron-dominated E to NE  rim of SN\,1006.  
We attribute the slower motion along most of the NW rim to the relatively recent encounter of the SNR 
shock with a higher density pre-shock sheet.   However, the existence of small regions with non-thermal   X-ray spectra and much higher shock speeds in close proximity to the more extensive regions with thermal spectra and slower shock speeds suggests low-density pockets within this denser sheet, or other small-scale inhomogeneities.

\acknowledgements

We acknowledge the conscientious support from members of the mission-planning staff at the \chandra\ X-ray center in planning our observations, as well as 
valuable comments on this paper by Una Hwang.
Primary support for this work is provided by the National Aeronautics and 
Space Administration through \chandra\ Grant Number GO2-13066, issued 
by the \chandra\ X-ray Observatory Center, which is operated by the 
Smithsonian Astrophysical Observatory for and on behalf of NASA under 
contract NAS8-03060\@.   SK is supported by the Special Postdoctoral Researchers Program in RIKEN, and PFW  acknowledges additional support from 
the National Science Foundation through grant AST-0908566.  

\bibliographystyle{apj}


\clearpage

\begin{center}
\begin{deluxetable}{rcccccc}
\tablecaption{\chandra\ ACIS-S Observations of SN\,1006, NW}
\tablewidth{0pt}

\tablehead{
 \colhead{ObsID} &
 \colhead{RA(J2000.)} &
 \colhead{Dec(J2000.)} &
 \colhead{Roll} &
 \colhead{Obs.~Date} &
 \colhead{Exposure} &
 \colhead{PI}
}
\scriptsize

\startdata
1959 &  15:02:22.5 &  -41:47:08 &  30.2$\degr$ &  2001 Apr 26  &  89.0 &  Long \\
13737 &  15:02:15.9 & -41:46:10 & 31.7\degr  & 2012 Apr 20 & 87.1 & Winkler
\enddata
 \label{tab:obsns}
 \end{deluxetable}
\end{center}

\begin{deluxetable}{lcccccccccc}
\tabletypesize{\footnotesize}
\rotate
\tablecaption{Positions of Fiducial X-ray Point Sources}
\tablewidth{0pt}

\tablehead{
\colhead{Source}&\multicolumn{3}{c}{2001} & & \multicolumn{3}{c}{2012} & & \multicolumn{2}{c}{Difference$^{b}$}\\ \cline{2-4}\cline{6-8}\cline{10-11}
& R.A. & Decl. & PSF size\tablenotemark{a} & & R.A. & Decl. & PSF size\tablenotemark{a} & & R.A. & Decl. 
}

\startdata
P1 & 15:02:07.010(0$\farcs$13) & -41:51:26.70(0$\farcs$15) & 1$\farcs$8 &
& 15:02:07.022(0$\farcs$12) & -41:51:26.13(0$\farcs$15) & 1$\farcs$9 &
& -0$\farcs$13 & 0$\farcs$57
\\
P2 & 15:02:30.468(0$\farcs$26) & -41:42:35.20(0$\farcs$36) & 1$\farcs$3 &
& 15:02:30.470(0$\farcs$20) & -41:42:35.12(0$\farcs$21) & 1$\farcs$2 &
& -0$\farcs$02 & 0$\farcs$08
\\
P3 & 15:02:34.514(0$\farcs$22) & -41:42:00.53(0$\farcs$18) & 1$\farcs$8 & 
& 15:02:34.54(0$\farcs$15) & -41:42:00.69(0$\farcs$09) & 1$\farcs$7 &
&  -0$\farcs$29 & -0$\farcs$16 
\\
P4 & 15:02:40.555(0$\farcs$13) & -41:45:09.02(0$\farcs$12) & 0$\farcs$8 &
& 15:02:40.495(0$\farcs$17) & -41:45:08.79(0$\farcs$13) & 1$\farcs$3 &
& 0$\farcs$67 & 0$\farcs$23 
\\
\hline 
\enddata
\tablenotetext{a}{Radius in which 39\% of total source photons are enclosed.}
\tablenotetext{b}{Positive values indicate apparent motions to the west and north in R.A.\ and Decl., respectively.}
\tablecomments{Values in parentheses are 1-$\sigma$ statistical uncertainties.}

\label{tab:points}
\end{deluxetable}

\begin{deluxetable}{lcccccc}
\tabletypesize{\footnotesize}
\tablecaption{Summary of Proper-Motion Measurements}
\tablewidth{0pt}
\tablehead{
\colhead{Region}&\colhead{Energy Band}&\colhead{Radius}&\colhead{Proper Motion\tablenotemark{a} }&\colhead{Expansion Index\tablenotemark{a}}&\colhead{Velocity\tablenotemark{a,b} }&\colhead{($\chi _\nu ^2$)$_{\rm min}$\tablenotemark{c} }  \\ &   (keV) & (\arcmin) & ($\asy$) & & ($\kms$) & }

\startdata
T1 & 0.5--1 & 14.1 & 0.32$\pm$0.03$\pm$0.03 & 0.38$\pm$0.03$\pm$0.04 & 3300$\pm$200$\pm$300 & 0.99 \\
T2 & 0.5--1 & 13.3 & 0.28$\pm$0.02$\pm$0.03 & 0.34$\pm$0.03$\pm$0.04 & 2800$\pm$200$\pm$300 & 0.58 \\
T2b & 0.5--0.6 & 13.3 & 0.28$\pm$0.03$\pm$0.03 & 0.35$\pm$0.04$\pm$0.04 & 2900$\pm$300$\pm$300 & 0.76 \\
T2c & 0.8--3 & 13.3 & 0.28$\pm$0.06$\pm$0.03 & 0.35$\pm$0.08$\pm$0.04 & 2900$\pm$600$\pm$300 & 0.94\\
T3 & 0.5--1 & 14 & 0.35$\pm$0.05$\pm$0.03 & 0.42$\pm$0.06$\pm$0.03 & 3700$\pm$500$\pm$300 & 0.84 \\
NT1 & 1--8 & 14.7 & 0.48$\pm$0.05$\pm$0.03 & 0.54$\pm$0.06$\pm$0.04 & 5000$\pm$500$\pm$300 & 1.32 \\
NT2 & 1--8 & 15.2 & 0.49$\pm$0.04$\pm$0.03 & 0.53$\pm$0.05$\pm$0.03 & 5000$\pm$400$\pm$200 & 1.13 \\
\hline 
\enddata 
\tablenotetext{a}{The first and second error terms  represent 1-$\sigma$ statistical and registration uncertainties, respectively.}
\tablenotetext{b}{Assumes a distance of 2.2 kpc.}
\tablenotetext{c}{Minimum $\chi^2$ per degree of freedom.}

\label{tab:prop}
\end{deluxetable}

\clearpage

\begin{figure}
\plotone{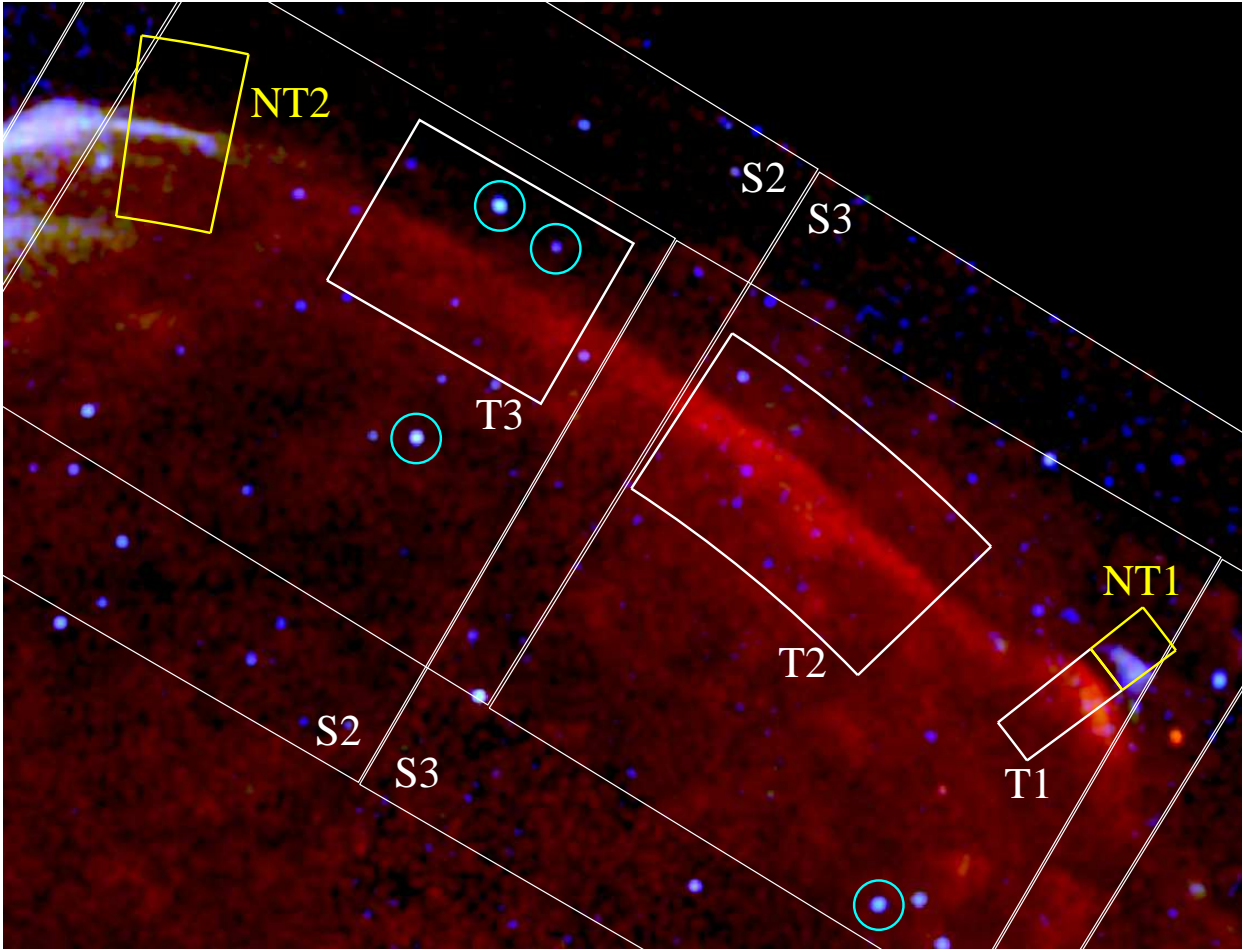}
\caption{{\it Chandra} three-color image of the northwestern limb of 
  SN\,1006.   Red, green, and blue correspond to fluxes in the 0.5--1\,keV, 1--2\,keV, 
  and 2--7\,keV bands, respectively.  The image is binned by 
  1\arcsec\  and has been smoothed by a Gaussian
  kernel of $\sigma = 6\arcsec$.  The intensity is scaled as the
  square root of the count rate.  The fields of view of the {\it Chandra} observations
  of the northwestern limb (ObsIDs 1959 and 13737 covering relatively inner 
  and outer regions, respectively) are shown as white boxes with ACIS
  chip identifications, and the four fiducial sources (P1 - P4 from west to east) used to register the images are indicated by the cyan circles .  The regions where we extract radial profiles are 
  shown in white (thermal dominated) and yellow (nonthermal dominated); note the spectral difference between the prominent features in the thermal vs nonthermal regions.  The field measures $17\arcmin \times 13\arcmin$ and is oriented north up, east left.
} 
\label{fig:image}
\end{figure}

\begin{figure}
\epsscale{0.8}
\plotone{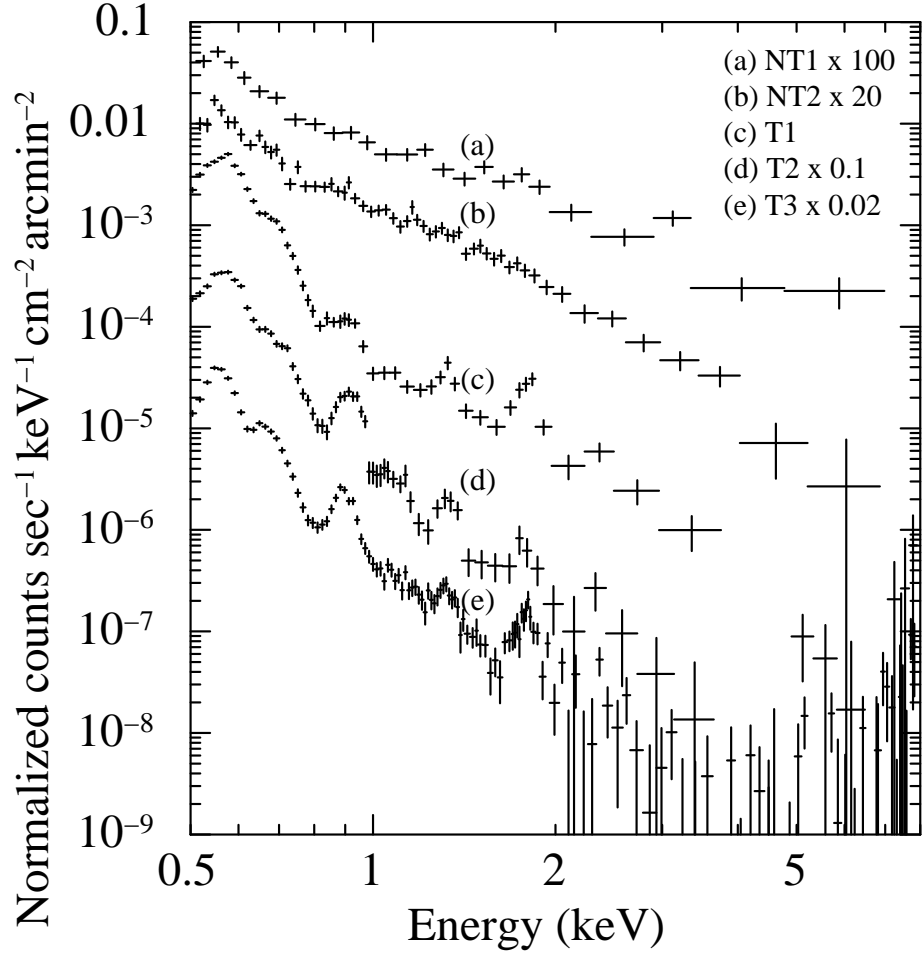}
\caption{{\it Chandra} ACIS spectra extracted from the thermal and non-thermal  regions indicated in
Fig.~1.  For clarity, spectra from different regions have been scaled by the factors indicated.  
The spectra have been extracted   not the entire the regions shown in Fig.~1, but instead from  the filamentary (T2, T3, and NT2) or knotty (T1 and NT1) features that we used for the proper-motion measurement.
} 
\label{fig:spectra}
\end{figure}

\begin{figure}
\epsscale{1.0}
\plotone{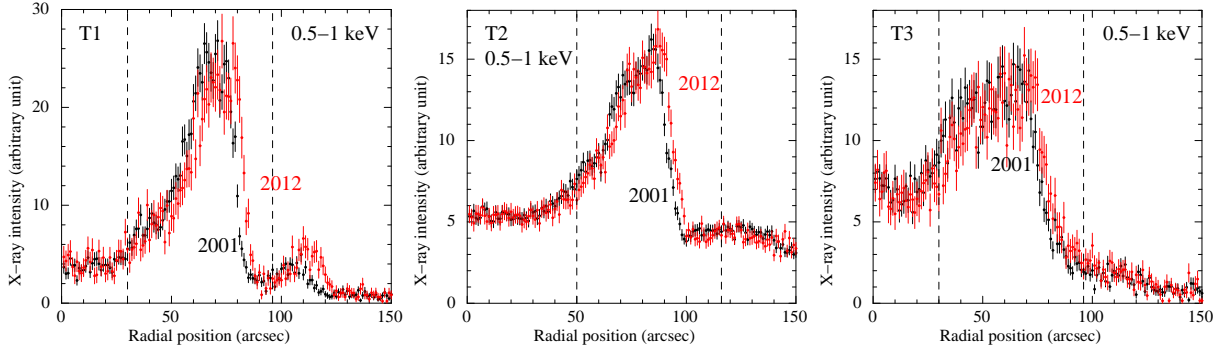}
\caption{Radial profiles extracted from the thermally-dominated areas
shown in Figure~\ref{fig:image}.  Data points in black and red represent 
the 2001 and 2012 epochs, respectively.  The intensity in 2012 is 
scaled to equalize that in 2001 by factors of 0.93, 1.01, and 1.11 
for T1, T2, and T3, respectively.  
The dashed vertical lines demarcate the regions used in the $\chi^2$ measurement of the proper motions.
The  profile labeled T1 (left panel) covers
both the T1 and NT1 regions in Figure~\ref{fig:image}; the smaller peak represents emission from the NT1 region, which is expanding faster than the thermal-dominated T1 peak.  Figure~\ref{fig:rad_profNT} (left) shows the profile from exactly this same area, but for higher energy X-rays.  
} 
\label{fig:rad_profT}
\end{figure}

\begin{figure}
\plotone{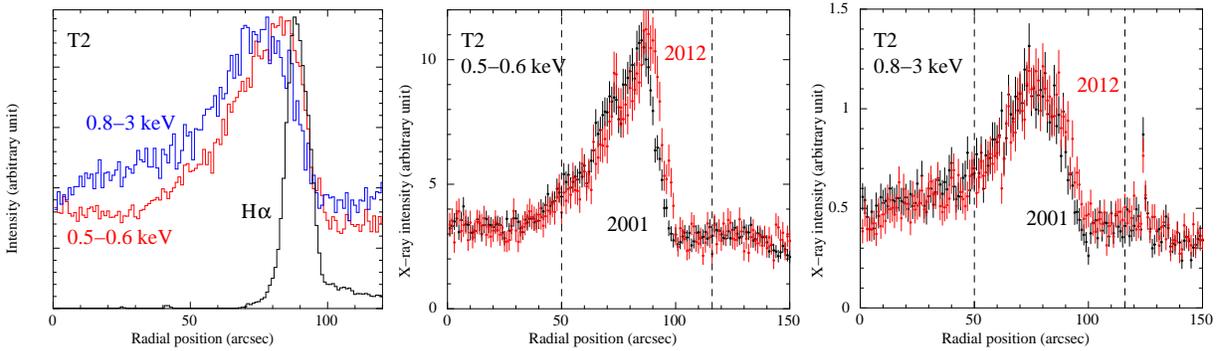}
\caption{Left: Radial profiles in the T2 region.  Black, red, and blue 
represent H$\alpha$, 0.5--0.6\,keV, and 0.8--3\,keV emission, 
respectively.  The X-ray data taken in the two epochs are combined 
after taking account of proper motions which amounts to 3 arcsec 
between the two observations.  The H$\alpha$ profile is shifted by an 
expected proper motion \citep{winkler03} at the X-ray observation.
Center and right: Same as Figure~\ref{fig:rad_profT}, but for different 
energy bands.  The intensity in 2012 is scaled to equalize that in 2001 by 
factors of 1.01 and 0.93 for 0.5--0.6\,keV and 0.8--3\,keV, respectively.
} 
\label{fig:rad_profT2}
\end{figure}

\begin{figure}
\plotone{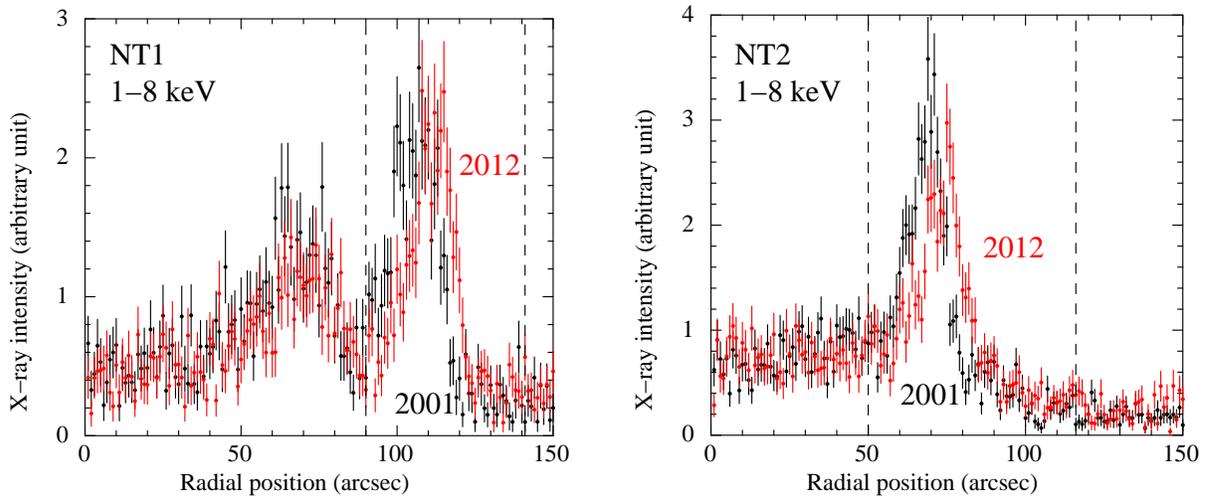}
\caption{Same as Figure~\ref{fig:rad_profT}, but for the NT1 and NT2 
regions, using the 1--8\,keV band.  The intensity of 2012 is scaled to 
equalize that of 2001 by factors of 1.09 and 0.94 for NT1 and NT2, 
respectively.  The  profile labeled NT1 (left panel) is from exactly the same area as the one labeled T1 in Figure~\ref{fig:image}; above 1 keV, emission from the outer, non-thermal region (NT1) is stronger than that from the inner, thermal one (T1).
} 
\label{fig:rad_profNT}
\end{figure}

\end{document}